\documentclass{article}

\usepackage{arxiv}

\usepackage[utf8]{inputenc} 
\usepackage[T1]{fontenc}    
\usepackage{hyperref}       
\usepackage{url}            
\usepackage{booktabs}       
\usepackage{amsfonts}       
\usepackage{nicefrac}       
\usepackage{microtype}      
\usepackage{cleveref}       
\usepackage{lipsum}         
\usepackage{graphicx}
\usepackage{natbib}
\usepackage{doi}

\title{OMP-Engineer: Bridging Syntax Analysis and In-Context Learning for Efficient Automated OpenMP Parallelization}

\author{ \href{https://orcid.org/0000-0000-0000-0000}{\includegraphics[scale=0.06]{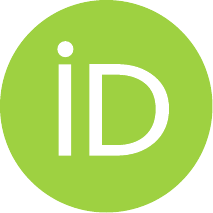}\hspace{1mm}Weidong Wang} \\
	Department of Software Engineering \\
    Faculty of Information Technology \\
	Beijing University of Technology \\
	\texttt{wangweidong@bjut.edu.cn} \\
	\And
	\href{https://orcid.org/0000-0000-0000-0000}{\includegraphics[scale=0.06]{orcid.pdf}\hspace{1mm}Haoran Zhu} \\
	Department of Software Engineering\\
    Faculty of Information Technology \\
	Beijing University of Technology\\
	\texttt{S202375042@emails.bjut.edu.cn} \\
}

\renewcommand{\shorttitle}{\textit{arXiv}}

\hypersetup{
pdftitle={OMP-Engineer},
pdfsubject={},
pdfauthor={Weidong Wang, Haoran Zhu},
pdfkeywords={OpenMP, Parallel Computing, Automation},
}

\begin{document}
\maketitle

\begin{abstract}
In advancing parallel programming, particularly with OpenMP, the shift towards NLP-based methods marks a significant innovation beyond traditional S2S tools like Autopar and Cetus. These NLP approaches train on extensive datasets of examples to efficiently generate optimized parallel code, streamlining the development process. This method's strength lies in its ability to swiftly produce parallelized code that runs efficiently. However, this reliance on NLP models, without direct code analysis, can introduce inaccuracies, as these models might not fully grasp the nuanced semantics of the code they parallelize. We build OMP-Engineer, which balances the efficiency and scalability of NLP models with the accuracy and reliability of traditional methods, aiming to enhance the performance of automating parallelization while navigating its inherent challenges.
\end{abstract}


\section{Introduction}
In the contemporary landscape of computing, parallel programming stands as a pivotal technology, crucial for enhancing computational efficiency and capitalizing on the capabilities of modern multi-core systems. The ability to concurrently execute multiple computations not only significantly reduces processing times but also optimizes the utilization of computational resources \cite{intro1, intro2}. Despite its advantages, the manual parallelization of code presents a formidable challenge—marked by its time-consuming nature and a high propensity for errors. These challenges stem from the intricate demands of identifying segments of code suitable for parallel execution, ensuring data consistency, and managing thread synchronization, among others. The complexity and expertise required to manually parallelize code have thus fostered the development of automated parallelization techniques, with two primary methodologies emerging to the forefront: Source-to-Source (S2S) conversion and Natural Language Processing (NLP)-based methods.

\subsection{Source-to-Source Tools}
At the heart of S2S approaches lies the principle of transforming sequential code into its parallel counterpart through automated syntactic modifications. These tools, grounded in the analysis of code structure and syntax, aim to streamline the parallelization process by identifying parallelizable code segments and applying standardized parallelization patterns. The strengths of S2S methods are manifest in their precision and the reduced likelihood of introducing errors during the parallelization process, attributed to their deterministic nature and reliance on established programming paradigms. However, the resultant parallel code often suffers from drawbacks such as diminished readability and lower execution efficiency. This is because S2S transformations tend to prioritize correctness over optimization, leading to parallelized code that, while functional, may not be optimized for performance. Additionally, the evolution of parallel programming standards, exemplified by the continuous updates to OpenMP, poses a significant challenge for S2S tools. Their inability to rapidly adapt to these evolving standards means that they can quickly become outdated, limiting their effectiveness and applicability in modern parallel programming projects.

\subsection{NLP-based Tools}
In contrast, NLP-based parallelization techniques represent a more flexible approach, leveraging the advancements in machine learning and artificial intelligence. By analyzing vast datasets of code, these methods train models capable of understanding both the structure and intent behind code snippets, thus generating parallel code that is not just syntactically correct but also optimized for performance. The primary advantage of NLP methods lies in their adaptability and the high execution efficiency of the generated code. These approaches are inherently designed to evolve alongside programming standards like OpenMP, ensuring their long-term relevance and utility. Furthermore, the broad applicability of NLP techniques across various programming scenarios and their ability to produce code that aligns with the latest parallelization paradigms underscore their potential as a transformative force in parallel programming. However, the reliance on learned patterns without a deep, semantic understanding of the original code introduces a notable drawback: the risk of errors. The generated parallel code, while generally efficient, may occasionally misinterpret the original code's intent or overlook complex dependencies, leading to inaccuracies or suboptimal parallelization outcomes.

\subsection{OMP-Engineer: A Hybrid Approach}
Recognizing the limitations inherent in both S2S and NLP methods, we introduce OMP-Engineer, a novel tool designed to bridge the gap between these approaches, harnessing their strengths while mitigating their weaknesses. OMP-Engineer is developed with the aim of providing a robust, adaptable, and error-resilient solution for the parallelization of code. By integrating the deterministic, syntax-based analyzation capabilities of S2S methods with the versatility of NLP techniques, OMP-Engineer offers a comprehensive approach to automated parallelization.

Our method commences with the generation of the Abstract Syntax Tree (AST) for the input code, which facilitates a deep syntactic analysis. This step ensures the parallelization process's accuracy by accurately identifying code segments that can be parallelized without error. Following the syntactic analysis, OMP-Engineer employs in-context learning to instruct Large Language Models (LLMs) on utilizing OpenMP directives efficiently and correctly. This hybrid approach enables the system to not only understand the structure of the code but also to grasp its semantic nuances, allowing for the generation of parallel code that is both precise and optimized for performance.

\section{Method}
\subsection{Syntax Analysis}
The methodological foundation of OMP-Engineer begins with a detailed syntax analysis, a critical step in ensuring the accuracy and integrity of the parallelization process. This analysis is initiated by leveraging ANTLR (Another Tool for Language Recognition) to construct an Abstract Syntax Tree (AST) from the source code. The AST serves as a comprehensive representation of the program's structure.

Following the generation of the AST, our approach delves into a nuanced examination of the code's syntax. The first phase of this examination focuses on identifying statements that are inherently resistant to parallelization, such as I/O operations like ``printf()''. These operations, due to their sequential nature, are flagged as non-parallelizable to preserve the semantic integrity of the program. The second phase of the analysis is dedicated to scrutinizing the access patterns of variables within the code. By meticulously tracking how variables are read and written---identifying sequences of access and potential data dependencies---we can detect the presence of dependencies that may impede parallel execution.

\subsection{Educating the LLM through In Context Learning}
Upon successfully identifying code segments that are amenable to parallelization through our syntax analysis phase, we proceed to the second core component of our methodology: educating the Large Language Model (LLM) for optimal parallelization.

To achieve this, we introduce a set of In Context Learning (ICL) materials specifically designed to guide the LLM in understanding and applying OpenMP directives effectively. These materials consist of a curated collection of code examples that demonstrate various parallelization patterns, accompanied by detailed explanations of each pattern's rationale, applicability, and implementation nuances. By embedding these examples and explanations directly into the LLM's context, we provide it with a rich, focused learning environment that significantly enhances its ability to generate correct and efficient parallel code.

\section{Results}
\begin{table}[ht]
\centering
\caption{Execution time comparison}
\label{tab:compare}
\begin{tabular*}{0.7\textwidth}{c@{\extracolsep{\fill}} c@{\extracolsep{\fill}} c@{\extracolsep{\fill}} c@{\extracolsep{\fill}} c@{\extracolsep{\fill}} c@{\extracolsep{\fill}}}
\hline
    &              & BT             & LU             & SP             &    \\ \hline
    & Sequential   & 122.93         & 102.00         & 83.44          &    \\
    & OMP-Engineer & \textbf{52.47} & \textbf{56.81} & \textbf{44.18} &    \\
    & Autopar      & 510.42         & 361.55         & 82.23          &    \\
    & Par4all      & 71.01          & 380.06         & 52.19          &    \\ \hline
\end{tabular*}
\end{table}
In our comparative performance evaluation, OMP-Engineer was benchmarked against two established parallelization tools, Autopar \cite{autopar} and Par4all \cite{par4all}, using three widely recognized benchmarks: BT, LU, and SP (all of which are from the NAS benchmark \cite{nas}). These benchmarks are instrumental in evaluating the efficiency and effectiveness of parallel computing solutions. The performance metrics obtained from these benchmarks are great for demonstrating the superior capability of OMP-Engineer in optimizing code parallelization.

Our results in \Cref{tab:compare} clearly showcase OMP-Engineer's outstanding performance in reducing execution time across all benchmarks when compared to the sequential (non-parallelized) baseline and the other tools. 

The comparison with the sequential execution times illustrates the substantial performance gains achieved through the application of OMP-Engineer's parallelization methodologies. These gains underscore the tool's effectiveness in reducing program runtime, thereby enhancing computational efficiency.

It is important to note that despite considering including ChatGPT-generated parallel code in our evaluation, we encountered excessive inaccuracies in the parallel code produced by ChatGPT. The high error rate in the parallel code generated by ChatGPT rendered it infeasible for a fair and meaningful speed comparison in this context. This decision underscores the challenges inherent in automated code parallelization and highlights the significance of our approach with OMP-Engineer, which effectively addresses these challenges through a combination of syntax analysis and in-context learning for educating the LLM. Consequently, OMP-Engineer not only ensures the accuracy of parallelized code but also significantly enhances its execution efficiency, as evidenced by our benchmark results.

\section{Conclution and Future Work}
In conclusion, our study presents OMP-Engineer, a novel tool designed to optimize code parallelization through a sophisticated integration of syntax analysis and in-context learning for Large Language Models. The benchmark results unequivocally demonstrate OMP-Engineer's superior performance in reducing execution times across various tests when compared to existing tools such as Autopar and Par4all. The innovative approach of educating LLMs with specific in-context learning materials has proven to be highly effective, enabling the generation of not only syntactically correct but also performance-optimized parallel code.

Despite the promising results, there remains ample scope for future work to further refine and enhance the capabilities of OMP-Engineer. Future directions could include:

\begin{itemize}
  \item Expanding the range of parallelization patterns and directives covered by the in-context learning materials to encompass a wider variety of programming scenarios and challenges.
  \item Enhancing the scalability of OMP-Engineer to support larger and more complex codebases, including those used in high-performance computing (HPC) environments.
\end{itemize}

Our work lays a solid foundation for the future of automated code parallelization, offering promising avenues for both research and practical applications. As computational demands continue to escalate, OMP-Engineer may play an increasingly critical role in enabling developers to harness the full potential of parallel computing technologies efficiently and effectively.

\bibliographystyle{unsrtnat}
\bibliography{references}

\begin{thebibliography}{5}
\providecommand{\natexlab}[1]{#1}
\providecommand{\url}[1]{\texttt{#1}}
\expandafter\ifx\csname urlstyle\endcsname\relax
  \providecommand{\doi}[1]{doi: #1}\else
  \providecommand{\doi}{doi: \begingroup \urlstyle{rm}\Url}\fi

\bibitem[Andrade et~al.(2023)Andrade, Griebler, Santos, and Fernandes]{intro1}
Gabriella Andrade, Dalvan Griebler, Rodrigo Santos, and Luiz~Gustavo Fernandes.
\newblock A parallel programming assessment for stream processing applications
  on multi-core systems.
\newblock \emph{Computer Standards \& Interfaces}, 84:\penalty0 103691, 2023.

\bibitem[Zeng(2023)]{intro2}
Guang Zeng.
\newblock Performance analysis of parallel programming models for c++.
\newblock In \emph{Journal of Physics: Conference Series}, volume 2646, page
  012027. IOP Publishing, 2023.

\bibitem[Liao et~al.(2009)Liao, Quinlan, Willcock, and Panas]{autopar}
Chunhua Liao, Daniel~J. Quinlan, Jeremiah~J. Willcock, and Thomas Panas.
\newblock Extending automatic parallelization to optimize high-level
  abstractions for multicore.
\newblock In Matthias~S. M{\"u}ller, Bronis~R. de~Supinski, and Barbara~M.
  Chapman, editors, \emph{Evolving OpenMP in an Age of Extreme Parallelism},
  pages 28--41, Berlin, Heidelberg, 2009. Springer Berlin Heidelberg.
\newblock ISBN 978-3-642-02303-3.

\bibitem[Amini et~al.(2012)Amini, Creusillet, Even, Keryell, Goubier, Guelton,
  Mcmahon, Pasquier, P{\'e}an, and Villalon]{par4all}
Mehdi Amini, B{\'e}atrice Creusillet, St{\'e}phanie Even, Ronan Keryell, Onig
  Goubier, Serge Guelton, Janice~Onanian Mcmahon, Fran{\c c}ois-Xavier
  Pasquier, Gr{\'e}goire P{\'e}an, and Pierre Villalon.
\newblock {Par4All: From Convex Array Regions to Heterogeneous Computing}.
\newblock In \emph{{IMPACT 2012 : Second International Workshop on Polyhedral
  Compilation Techniques HiPEAC 2012}}, Paris, France, January 2012.
\newblock 2 pages.

\bibitem[Bailey et~al.(1991)Bailey, Barszcz, Barton, Browning, Carter, Dagum,
  Fatoohi, Frederickson, Lasinski, Schreiber, et~al.]{nas}
David~H Bailey, Eric Barszcz, John~T Barton, David~S Browning, Robert~L Carter,
  Leonardo Dagum, Rod~A Fatoohi, Paul~O Frederickson, Thomas~A Lasinski, Rob~S
  Schreiber, et~al.
\newblock The nas parallel benchmarks—summary and preliminary results.
\newblock In \emph{Proceedings of the 1991 ACM/IEEE Conference on
  Supercomputing}, pages 158--165, 1991.

\end{thebibliography}

\end{document}